\pdfoutput=1

\documentclass[10pt, conference, compsocconf]{IEEEtran}

\usepackage{ifpdf}

\ifCLASSINFOpdf
\else
\fi

\newtheorem{defn}{{\bf{Definition}}}

\usepackage{graphicx}
\usepackage{mathtools}
\usepackage{hhline}
\usepackage[utf8]{inputenc}
\usepackage{etoolbox}

\usepackage{amsmath,amssymb,amsfonts}

\usepackage{url}
\usepackage{verbatim}
\usepackage{multirow}
\usepackage{caption}
\usepackage{subcaption}

\usepackage[linesnumbered,ruled,vlined]{algorithm2e}
\usepackage{algpseudocode}
\usepackage{mathtools}

\usepackage{ifpdf}
\usepackage{lmodern}
\usepackage{lipsum}
\usepackage[T1]{fontenc}
\usepackage{textcomp}
\usepackage{mdwlist}
\usepackage{pbox}
\usepackage{comment}
\usepackage{ifthen}
\usepackage{color}
\usepackage{colortbl,xcolor}
\usepackage{upgreek}
\usepackage{upgreek}
\usepackage{todonotes}
\usepackage{wasysym}

\usepackage{tikz}
\usetikzlibrary{arrows}

\usepackage[hidelinks,bookmarks=false]{hyperref}

\definecolor{Gray}{gray}{0.9}
\definecolor{LightCyan}{rgb}{0.88,1,1}

\definecolor{LightCyan}{rgb}{0.88,1,1}
\definecolor{Gray}{gray}{0.9}
\definecolor{light-gray}{gray}{0.95}

\begin{document}
%
\title{Klout Score: Measuring Influence Across Multiple Social Networks}


\author{\IEEEauthorblockN{Adithya Rao,
Nemanja Spasojevic,
Zhisheng Li and 
Trevor DSouza}
\IEEEauthorblockA{Lithium Technologies | Klout\\
San Francisco, CA\\
Email: adithya, nemanja, zhisheng.li, trevor.dsouza@klout.com}
}

\maketitle

\begin{abstract}
\label{section:abstract}
In this work, we present the Klout Score, an influence scoring system that assigns scores to 750 million users across 9 different social networks on a daily basis.
We propose a hierarchical framework for generating an influence score for each user, by incorporating information for the user from multiple networks and communities. 
Over 3600 features that capture signals of influential interactions are aggregated across multiple dimensions for each user.
The features are scalably generated by processing over 45 billion interactions from social networks every day, as well as by incorporating factors that indicate real world influence. 
Supervised models trained from labeled data determine the weights for features, and the final Klout Score is obtained by hierarchically combining communities and networks.
We validate the correctness of the score by showing that users with higher scores are able to spread information more effectively in a network.
Finally, we use several comparisons to other ranking systems to show that highly influential and recognizable users across different domains have high Klout scores.
\end{abstract}


%
\IEEEpeerreviewmaketitle

\section{Introduction}
\label{section:introduction}
It is estimated that there are now over a billion users on online social networks, exceeding even the number of websites on the internet.
In the past decade, ranking webpages based on importance of linked content, clicks and impressions led to ubiquitous internet applications such as search.
Applying effective ranking techniques to determine influential users on the internet has a similar potential to lead to many new and useful applications as well.

When a user posts a message on social media, other users in the network who see the content may perform certain actions in reaction to the original message.
The fact that the original message prompted certain reactions from other users is an indication that the user influenced them in some manner.  
For example, a user may post a message on Facebook about her experience in a restaurant, with a link to the restaurant's webpage.
A user who reads the original message may choose to react to it in several ways such as: read the message, click on the link to get more information, reshare the link with other users in his own network, or actually visit the restaurant for dinner.
The type of reaction gives an indication of the strength of influence the message had on the user.

There are, of course, many variables pertaining to offline actions that cannot be directly  measured, such as the effect of seeing a billboard on a freeway. 
But in the context of social media, a large set of user reactions such as impressions, clicks, likes, comments, reshares and purchase behavior is measurable.
By observing the quantity and quality of reactions that a user generates among other users in the network, it is therefore possible to get a measure of how influential he or she is.

Here we introduce the \textit{Klout Score} as a metric for measuring influence of users on online platforms such as social networks and community forums. 
While the Klout Score has been available since 2008, early versions of the score included fewer signals, and were therefore less effective as a metric of influence \cite{anger2011klout}.
However, because the system was built to be extensible and flexible, the Klout score has evolved to incorporate many new sources of information, growing more accurate over time. 
Today, the Klout score is widely used for identifying influential users for applications such as targeted search and influencer marketing \cite{schaefer2012influence}.

It is this extensible and flexible framework that we present in this study, along with results that validate the effectiveness of the score in measuring influence. Our contributions in this paper are as below:
\begin{itemize}
  \item\textbf{Scalable Production System:} We describe a full production system that assigns \textit{Klout Scores} to 750 million public and registered user profiles from 9 different networks, by processing 45 billion interactions everyday. 
  \item\textbf{Feature Generation:} We outline how features that capture different aspects of influential actions are generated. In our models we use over 3600 such features.
  \item\textbf{Hierarchical Scoring:} We explain how networks are scored individually, and are then combined into a single score using a hierarchical approach.
  \item\textbf{Validation:} We present experiments and comparisons that show that the Klout score effectively measures influence in a variety of contexts.
\end{itemize}

\section{Problem Setting}
\label{section:problem_setting}
\subsection{Related Work}
Recently, a great deal of research work has emerged on exploring the social influence measurement and applications.
Tang et al. \cite{tang2009social} analyzed topic-based social influence on academic collaboration networks.
The authors of \cite{agarwal2008identifying, weng2010twitterrank} identified influencers on blogs and the Twitter social network respectively.
In comparison, here we consider influence simultaneously on multiple networks.

Influence maximization \cite{Kempe2003MSI, Chen2009EIM, chen2015online} is the problem of finding a subset of nodes that would maximize the spread of information in a given graph.
This problem differs from that of assigning influence scores to every node in the graph, since the former is a targeting or subset selection problem, while the latter is a measurement or ranking problem.

Several metrics have been used in previous work to measure influence.
Alexy et al. \cite{khrabrov2010discovering} used a PageRank-like social interaction score and number of mentions over time to measure user influence on Twitter.
Behnam et al. \cite{hajian2011modelling} modeled influence using metrics such as number of followers and ratio of affection.
Influence measurement on social networks also has a temporal aspect to it, since messages typically have short lifespans in terms of recieving reactions \cite{Spasojevic:when-to-post}. 
This problem of understanding time-sensitive influence is one that has remained relatively under-explored in previous work.
Here we present a framework for feature engineering that allows granular measurement across various dimensions and scales to thousands of features.

The Klout score has been applied to various marketing applications \cite{schaefer2012influence}, and has been used in studies about social behavior \cite{edwards2013credibility}.
The field of studying influence is still in its early stages, and this work aims at advancing such influence measurement systems.

\subsection{Problem Statement}

While \textit{influence} is a broad and subjective concept, we can quantitatively describe it in terms of observable reactions to stimuli.
If an entity performed an observable reaction in response to a stimulus originating from another, then we can say that the latter influenced the former in some manner.
Thus we can consider the influence of an entity to be the ability to induce reactions in other entities.

More specifically, in the context of social networks, we can define the influence of a user to be the ability to induce reactions in other users. 
Thus an influence score determines how effectively a user may be able to influence other users via his or her actions. 

Let $\mathcal{G}$ represent a network or community of users, who interact with each other via a set of actions $\mathcal{A}$. 
An influence score can then be defined as:
\begin{defn}\label{def:influence_score}
{\bf{Influence Score:}} For each user $u$ in a network $\mathcal{G}$, let $\mathcal{G}_u$ be the subset of the network containing the users who may directly or indirectly interact with $u$, via a set of reactions $\mathcal{R} \subseteq \mathcal{A}$.
Then an \textbf{influence score} $I(u, T)$ is a measure of the degree and quantity of reactions that $u$ can induce in $\mathcal{G}_u$ over a specified time period $T$.
\end{defn}

The score thus determined can be used to relatively compare influential users in the network.
Below we describe a few of the factos that an influence scoring system needs to take into account.

\subsection{Considerations}
There are several aspects that an influence scoring system must incorporate in order to be effective:

\paragraph{User Scalability}
An effective influence scoring system must be able to process information from the complete network graph, which may include hundreds of millions of users.  
Some previously suggested approaches rely on loading the entire graph in-memory to perform such computation, but these approaches have limitations when dealing with web-scale datasets. 
We solve this problem by leveraging batch processing frameworks that aggregate and generate features for each user separately in multiple passes, before combining them hierarchically.

\paragraph{Network Scalability}
A user's online persona typically spans multiple social and professional networks.
An influence scoring system must therefore be able to scale across these different networks, to unify the available information for a user.
It should also be able to handle the distinct sets of interactions and user behavior patterns associated with each network.
Further, the influence scoring system must also determine the relative importance of networks when they are combined together.
We discuss these aspects of influence scoring in the following sections.


\paragraph{Interaction Graph}
As shown in \cite{Kempe2003MSI} influence measurement strategies that relies solely on structural properties of the graph such as degree and centrality heuristics do not perform well, and it is essential to consider information dynamics in the network. 
Thus in addition to properties such as in-degree and centrality of nodes in a graph, an influence measurement strategy must identify and capture variables that indicate dynamic information flows.
Furthermore, the manner of interaction indicates the strength of influence, and some interactions may indicate a greater degree of influence than others.
This relative importance of interactions may be determined by constructing granular features that can be weighted individually.

\paragraph{Temporal factors}
The variables that capture influence may be broadly categorized as those that capture long-lasting influence, versus those that capture dynamic and changing influence.
Since the importance of dynamic variables fade over time \cite{Spasojevic:when-to-post}, an influence measurement system must be sensitive to time decay of influential interactions. 
In our system, we choose a time window of $90$ days to consider dynamic interaction behavior, in addition to signals that capture long-lasting influence outside this time window.

\paragraph{Offline factors}
It is plain that signals on social networks are only a partial representation of a user's overall influence, and can only provide limited accuracy for influence measurement.
It is therefore crucial to incorporate proxy sources that signify a user's real world influence. 
Here we use Wikipedia and news articles to extract signals that may indicate the user's offline influence.

\paragraph{Reach and Strength of Influence}
The size of $\mathcal{G}_u$ may vary widely for different users in the network, and an influence scoring system must determine the importance of the user's reach with respect to the number of reactions.
Further, the manner and frequency of reactions may determine how strongly a user influences another.
Thus a user who induces a total of 100 reactions among 10 other users may or may not have the same influence score as a user who induces 100 reactions among 50 users, depending on the strategy chosen for scoring.
An influence scoring system may choose to score the latter higher, since she reaches a larger set of people; while another may score the former higher, since he is able to more strongly influence a smaller set of users.
The chosen strategy may depend on the application. 

\section{System Overview}
\label{section:system_overview}
\subsection{Methodology}
Here we propose a hierarchical approach to compute influence scores. 
We build an interaction graph by capturing reactions generated in response to social media posts. 
The reaction types chosen are those that are strong indicators of influential information flows between nodes.
In addition we derive information from the relatively slow-changing graph structure as well.
To factor in temporal effects and time decay, a trailing window of activity over 90 days is used.
More recent actions have a greater significance compared to older actions, all other variables remaining the same.
Features such as PageRank derived from Wikipedia and number of news article mentions provide indicators of real world or offline influence. 

Features derived from such information are used to create feature vectors for each user, for each network or community. 
Supervised machine learning models are built using ground truth labels generated for each network.
The model weights applied to the network feature vector for a user gives a network score.
The overall score for a user is computed by combining the scores from all networks and communities where the user has a presence, in a hierarchical manner.

While no measurement system can claim to comprehensively capture all signals of influence, we design our system such that it is flexible enough to easily incorporate new information, as and when it becomes available. 

\subsection{Pipeline}
\begin{figure}
  \centering
  \fbox{\includegraphics[width=0.97\columnwidth]{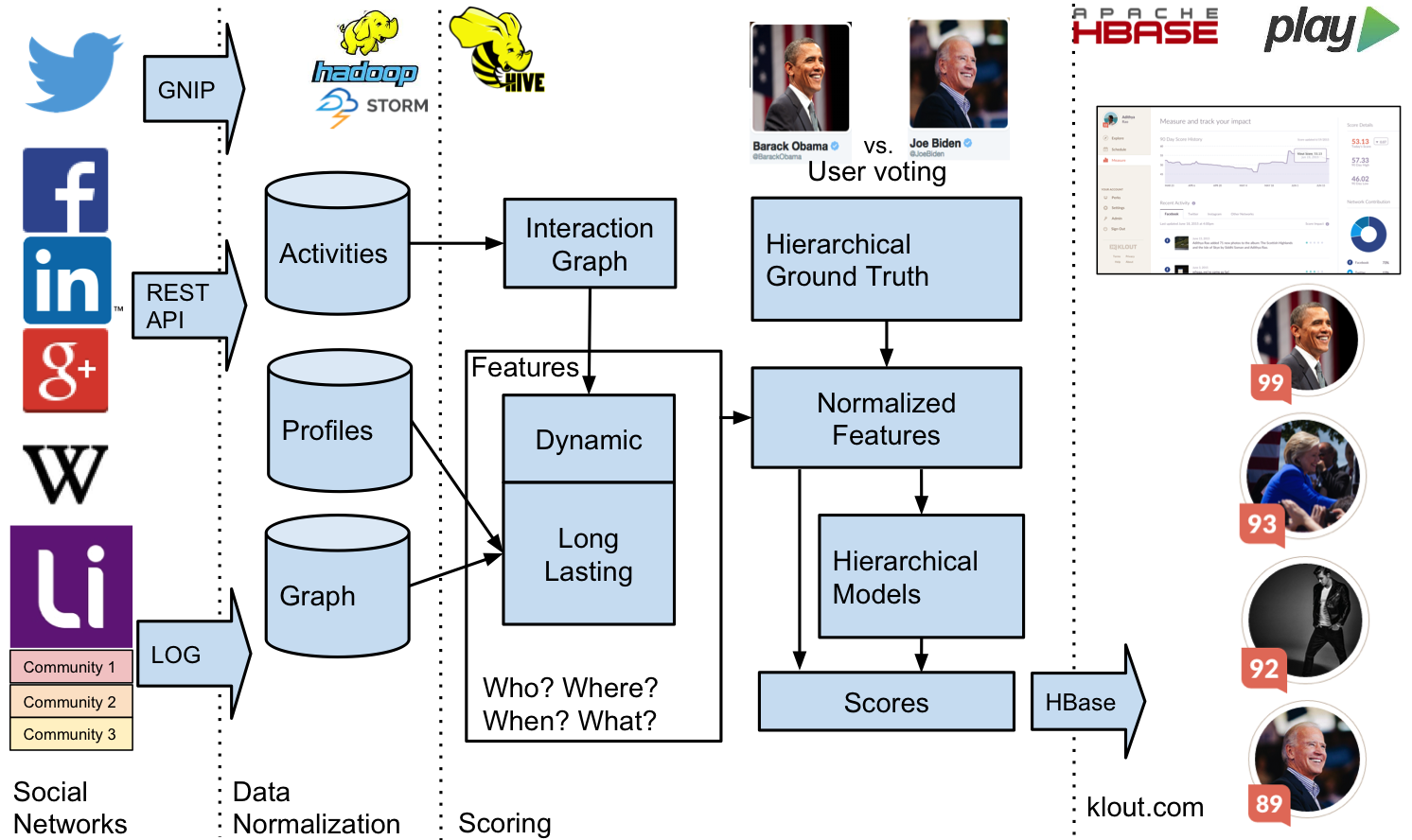}}
  \setlength{\abovecaptionskip}{0pt}
  \caption{Scoring Pipeline Overview}
  \label{figure:scoring_pipeline}
\end{figure}

Klout scores are computed for 750 million users from 9 major networks including Twitter (TW), Facebook (FB), LinkedIn (LI), Google+ (GP), Foursquare (FS), Instagram (IG), YouTube (YT) and Lithium Communities (LT), in addition to Wikipedia (WK).
When a user registers on Klout.com he associates his identities on different social networks with his Klout profile. 
For Twitter, public data is collected via the Mention Stream\footnote{https://gnip.com/sources/twitter/}; anonymized data for opted-in Lithium Communities\footnote{http://www.lithium.com/} comes from in-house datastores; and data for other social networks is collected via REST APIs on the user's behalf, based on the granted permissions.
All collected data is parsed and normalized to protocol buffers that encode user interactions, graph, and profile information. 
Data is collected continuously from interactions in a trailing window of 90 days using the Play Framework. 
The collected data is written out to a distributed file system, where batched parsing and processing is done using Hadoop MapReduce and Hive.
The batch processing pipeline derives features for each user, normalized against the global population.
Feature weights from offline models built using the ground truth data are then applied to generate Klout scores.
The pipeline overview is shown in Figure \ref{figure:scoring_pipeline}.
Over 45 billion interactions are processed in each pipeline run, with 0.5 billion new interactions added each day.
The daily footprint of the pipeline is 196.14CPU days, with 18.46TB of reads and 9.53TB writes.

\subsection{Features}
In order to build a supervised model for influence scores, we generate a set of quantitative features for each user who is represented by a node in the graph. 
In some cases such as LinkedIn Job Titles or Community Badges, the categorical variables are converted to quantitative values based on the ordered list of the categories.
Thus all features are designed to be directly proportional to influence. 

The features may be broadly divided into two types - long-lasting and dynamic.
Long-lasting features include those that change gradually or infrequently. 
Education history and Wikipedia PageRank are examples of such features. 
The types of long-lasting variables are summarized in Table \ref{long-lasting-types}, although this is not an exhaustive list.
Over 60 such long-lasting features are considered as part of the Klout score. 

\begin{table}[htdp]
\caption{Types of Long-lasting features}
\begin{tabular}{|p{1.5cm}|p{4.2cm}|p{1.6cm}|}
\hline
\rowcolor{Gray} \textbf{Feature Type} & \textbf{Features} & \textbf{Networks} \\
\hline
Node Degree & Followers, Friends, Fans, Subscribers, In-links & TW, FB, IG, GP, WK, YT \\ \hline
Graph Properties & {PageRank, Inlink to Outlink ratio} & WK \\ \hline
Categories & Job Title, Education Level, Endorsements, Recommendations, Awards, Community Badges & LI, LT \\
\hline
\end{tabular}
\label{long-lasting-types}
\end{table}%

\begin{table}[htdp]
\caption{Common Dimensions for Dynamic Features}
\begin{tabular}{|p{1.2cm}|p{1cm}|p{1.4cm}|p{1.0cm}|p{1.9cm}|}
\hline
\rowcolor{Gray} \textbf{Audience (Who)} & \textbf{Time (When)} & \textbf{Network (Where)} & \textbf{Type (What)} & \textbf{Action (How)} \\
\hline
\parbox{1cm}{All, Higher, Peers} & \parbox{1cm}{3, 7, 14, 21, 30, 60, 90} & \parbox{1.4cm}{TW, FB, LI, GP, FS, IG, LT, YT} & \parbox{1.0cm}{Message, Photo, Video} & \parbox{1.9cm}{Comment, Reply, Like (Upvote), Mention (Tag), Reshare (Retweet, via), View (Impression) } \\
\hline
\end{tabular}
\label{feature-tuples}
\end{table}

The dynamic features, on the other hand, capture information flowing through edges in the graph between users. 
As described in previous sections, the primary signal of an influential interaction is when an action from a user leads to reactions among other users.
Each of these reactions indicates a unit of information flow.
A reaction can be represented by a tuple of dimensions (\textit{Who}, \textit{When}, \textit{Where}, \textit{What}, \textit{How}) as below:
\begin{itemize}
  \item\textbf{Who:} The characteristics of the audience who reacted to the original post from the user.
  \item\textbf{When:} The difference between the current time and the time at which the reaction occured.
  \item\textbf{Where:} The social network on which the reaction was performed.
  \item\textbf{What:} The unit of original content or action on which the reaction was performed.
  \item\textbf{How:} The type of reaction.
\end{itemize}

The first step while generating features is to normalize all the reactions based on the above dimensions as (\textit{actor}, \textit{timestamp}, \textit{network}, \textit{original content type}, \textit{action}).
Features are generated by aggregating all reactions that are represented by the same tuple.
For example, all the reactions of the kind \textit{"comments from a user's peers received on Facebook Photo posts in the last 7 days"} are aggregated into a single feature represented by the tuple \{\textit{Peers}, \textit{7 days}, \textit{Facebook}, \textit{Photo}, \textit{Comment}\}.

This aggregation is achieved in a single pass through the dataset, by employing User Defined Functions (UDFs) such as \textit{conditional\_emit} and \textit{multiday\_sketch}, applied within Hive queries that are executed as MapReduce jobs.
We have open sourced these UDFs in a project named Brickhouse\footnote{https://github.com/klout/brickhouse}.

This feature generation framework has several advantages.
Firstly, it allows a large set of dynamic features to be generated for training.
Table \ref{feature-tuples} provides the list of the most commonly used dimensions for generating dynamic features -- 3 cohorts of users, 7 time windows, 8 networks \footnote{Since Wikipedia is not a conventional social network, we exclude it when generating dynamic features.}, 3 common content types, and 6 common types of reactions on content.
Note that all the listed content types and actions may not be present for every network, nor are the dimensional values restricted to those in Table \ref{feature-tuples}.
Each network may have its own unique content types and actions, leading to additional features.
Overall around 3550 dynamic features are generated by the system, using various combinations of the dimensional values.

Secondly, the framework allows easy extensibility in any of the dimensions. 
Thus adding features for a new action or a network becomes only a matter of identifying the specific dimensional values needed.

Thirdly, this approach also provides granularity while learning a supervised model that assigns weights to features.
By allowing weights to be assigned to specific tuple combinations, the models can be made sensitive to changes in each of the dimensions. 
For example, the weights assigned to features that represent a \textit{reshare} action may carry a higher weight than a \textit{like} action, all other dimensions being the same.
Similarly a \textit{reshare} action by a user whose is more influential than the user himself may have a higher weight than the same action from one of the user's peers.

Finally, aggregating interactions into such dimensions can be done using multiple passes through the dataset, which has a computation complexity of $O(n)$. 
This means that the entire graph need not be loaded in-memory to perform the computation, and can instead be performed efficiently on a distributed batch processing framework such as MapReduce or Hive.

\subsection{Hierarchical Scoring}

This hierarchical architecture allows for extensibility both in terms of depth (granularity of features) as well as breadth (number of sources). This enables the scoring mechanism to be sensitive to signals as well as scalable in terms of networks.

\begin{figure}
  \centering
  \fbox{\includegraphics[width=0.97\columnwidth]{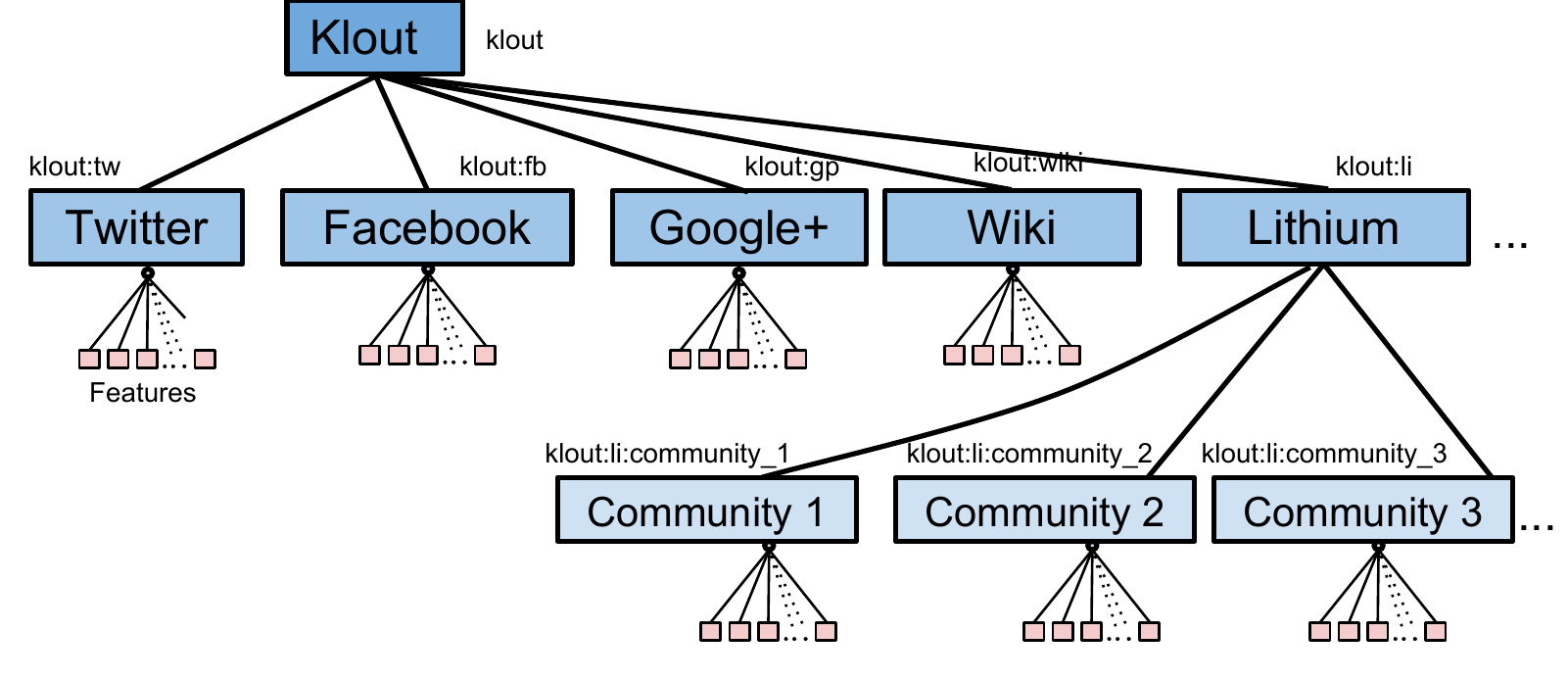}}
  \setlength{\abovecaptionskip}{0pt}
  \caption{Klout Score Structure Overview}
  \label{figure:score_composition_overview}
\end{figure}

Ground truth data is generated for individual networks and communities through tools designed to collect labeled data.
For generating this data, evaluators were shown pairs of users who were known to them on the network, and were asked to identify the more influential one in each pair.
Each pair of users was ranked by multiple users to reduce bias.
Close to 1 million such evaluations were collected for all networks combined, with the hundreds of thousands of labels for larger networks such as Twitter and Facebook, and tens of thousands for smaller networks.
To handle ambiguity in the labels, they were then pre-processed to only pick pairs with a clear winner by a difference of at least 2 votes.

The features generated typically have a power law distribution with a long tail. 
In order to make the feature values comparable, they are log normalized by the global maximum value per feature.
A feature vector with elements as these normalized values is then generated for each user.
Models are then trained for ranked users in the ground truth training set using supervised learning methods to generate a weight associated with each feature. 
Specifically, Non-Negative Least Squares (NNLS) regression is used for model building, since the features are designed such that an increase in a feature value corresponds to higher influence.
The trained models have an F1 score between $0.70$ to $0.75$ for most networks, which is relatively high given that human evaluators who provide the ground truth labels do not always agree on the ordering.

\begin{figure}
  \centering
  \fbox{\includegraphics[width=0.7\columnwidth]{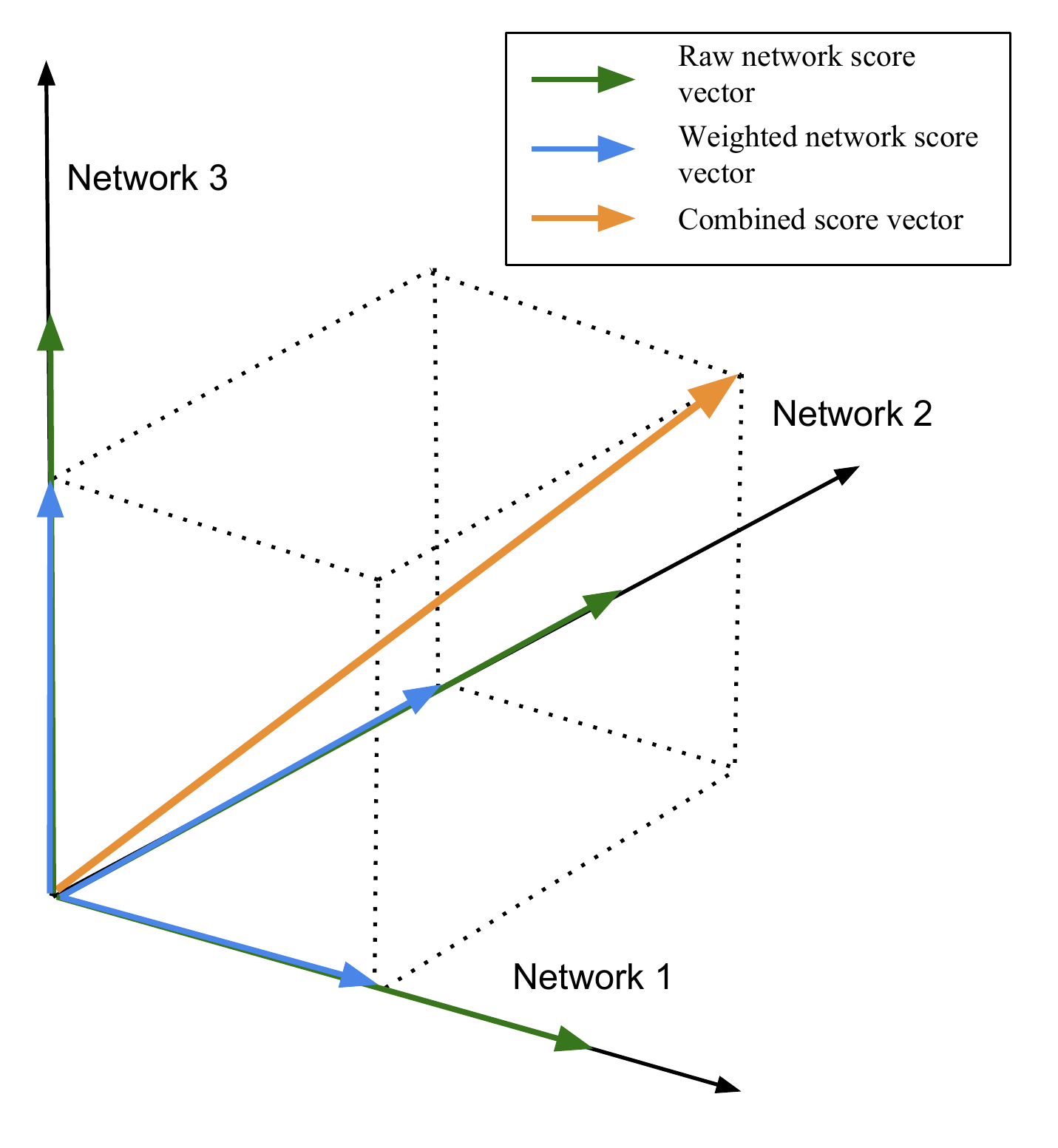}}
  \caption{Network Score Combiner (simplified to 3 network dimensions)}
  \label{figure:score_combiner}
\end{figure}

Let $n_{i,j}$ represent the $j^{th}$ network or community at the $i^{th}$ level in the hierarchy, with $i=0$ representing the topmost level.
For a user $u$ on a given network $n_{i,j}$, we denote the normalized feature vector as $\mathbf{f}(u, n_{i,j})$.
The weights associated with the learnt model for the network $n_{i,j}$ are given by the weight vector $\mathbf{w}(n_{i,j})$.
The application of these weights applied to a feature vector yields a score that is in the range of $[0,1]$.
Thus, for the graph $\mathcal{G}_u$ corresponding to the network $n_{i,j}$, and a time period $T$ over which the features are computed, the influence score for the user in that network is given by:
\begin{eqnarray}
 I_{i,j}(u, T) &=& s(u, n_{i,j}) \nonumber \\
               &=& {\mathbf{f}(u, n_{i,j})} \cdot {\mathbf{w}(n_{i,j})} \nonumber
\end{eqnarray}

The scores obtained for a user as a result of applying the learnt weights to the network or community feature vectors are further combined into a new vector for the next level in the hierarchy. 
Let the network $n_{i-1,K}$ on the level $i-1$ have $k$ different child networks corresponding in the $i^{th}$ level. 
Then the feature vector for $n_{i-1,K}$ is given by: 
\begin{equation}
{\mathbf{f}(u, n_{i-1,K})} = [ s(u, n_{i,1}), s(u, n_{i,2}), ... , s(u, N_{i,k})  ]
\end{equation}
The weight vector for this level can now be applied to get a score for $n_{i-1,K}$.

Thus, for networks with child communities, such as Lithium, the community scores for a user form a feature vector for the network level, which can be combined to get a network score. 
The network level scores are further combined at the root level to give a score that represents the influence of the user combined across the different networks where he is present.  

At higher levels in the hierarchy, it is challenging to get ground truth data that represent how networks or communities may be combined together. 
In the absence of such labeled data, it may not be possible to generate weight vectors using supervised models.
Instead, we extract weight vectors for the higher levels based on network or community graph properties. 
For instance, weights that represent the potential audience that a user on the network could influence may be derived from heuristics such as overall graph size or average node degree.

Further, unlike the features at the lower levels, the features at these higher levels may be fairly uncorrelated.
This allows us to approximate the child networks as different orthogonal axes to generate a vector space.
For such levels, the combined score may be computed as the Euclidean or L2 norm of the vector obtained by the component-wise product of the weight and network feature vectors. 
\begin{equation}
 s(u, n_{i,j}) = \lVert {\mathbf{f}(u, n_{i,j})} \ast {\mathbf{w}(n_{i,j})} \rVert
\end{equation}
where $\ast$ represents the operator for element-wise multiplication.

In particular, the raw Klout Score $KS_r(u)$, denoted by the network notation $n_{0,1}$ is given by the L2 norm of the network scores in level $i=1$:
\begin{eqnarray}
 KS_r(u) = I_{0,1}(u, T) &=& s(u, n_{0,1}) \nonumber \\ 
                         &=& \lVert {\mathbf{f}(u, n_{0,1})} \ast {\mathbf{w}(n_{0,1})} \rVert
\end{eqnarray}

This root level score is finally scaled to $[0,100]$, giving the Klout score.
Since the original features are log normalized, the final score is also interpreted to be on a logarithmic scale. 
Thus a user with a score of 60 may be $\alpha$ times as influential as a user with a score of 50, where $\alpha$ is the constant associated with the power law distribution.

In the next section, we perform validation on the Klout score via several comparisons.

\section{Validation}
\label{section:results}
This section examines the Klout Score from four different aspects to illustrate its correctness and usefulness.

\begin{figure}
  \centering
  \includegraphics[width=0.97\columnwidth]{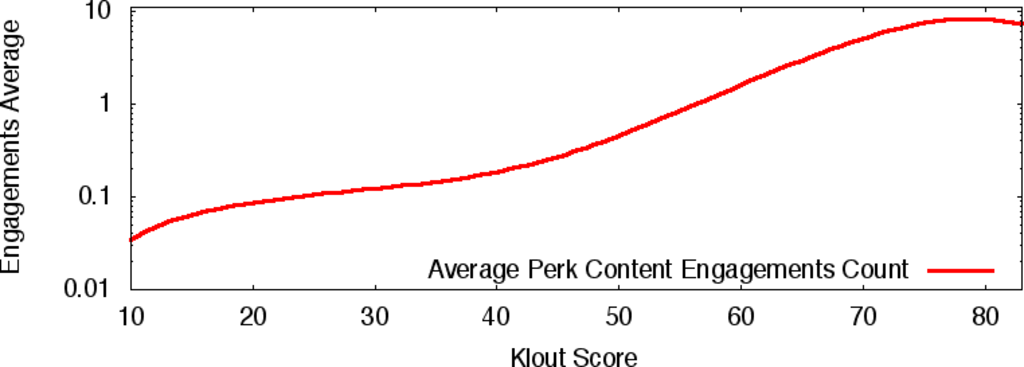}
  \setlength{\abovecaptionskip}{0pt}
  \caption{Analysis of average perk related content reaction count as a function of authors' average Klout score}
  \label{figure:perk_analysis}
\end{figure}

\begin{figure*}
\centering
\begin{subfigure}[b]{0.4\textwidth}
        \includegraphics[width=\textwidth]{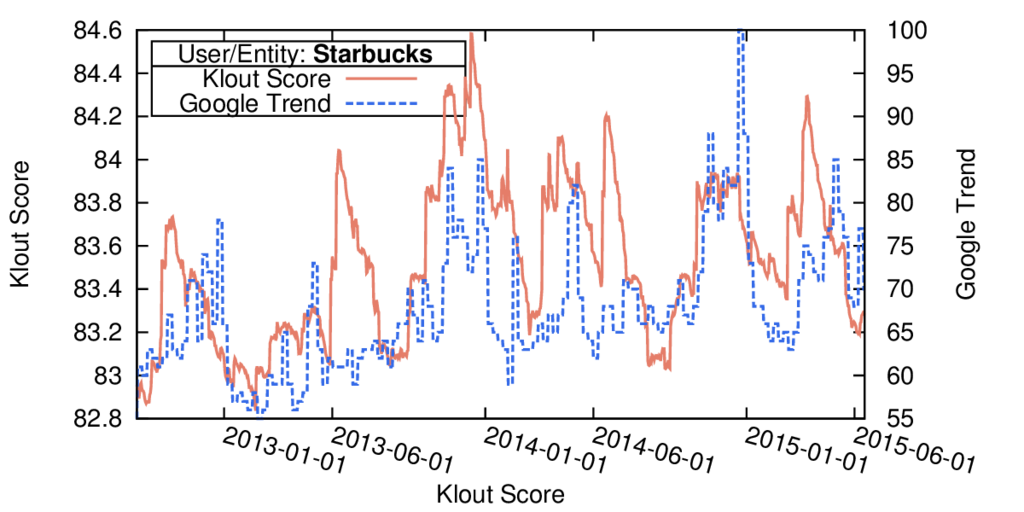}
\end{subfigure}
\begin{subfigure}[b]{0.4\textwidth}
        \includegraphics[width=\textwidth]{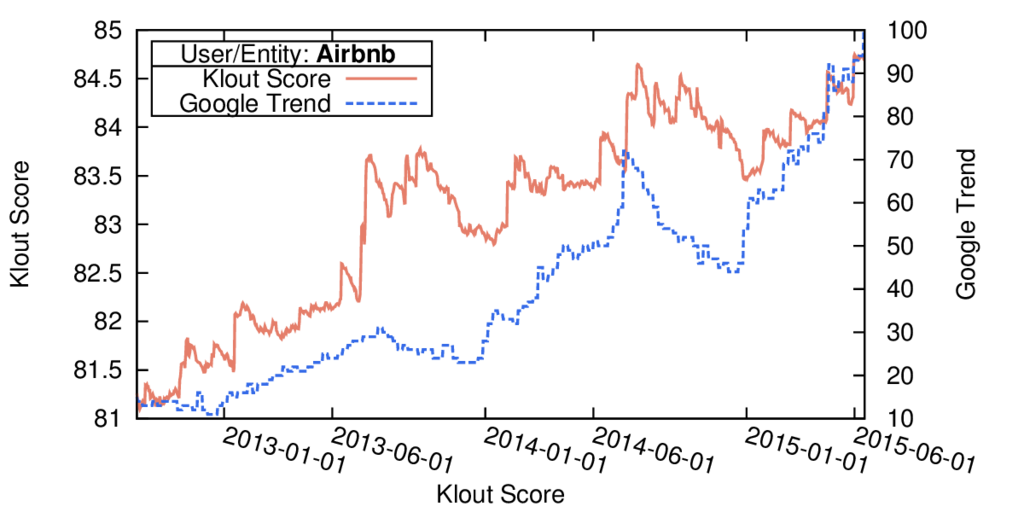}
\end{subfigure}
\begin{subfigure}[b]{0.4\textwidth}
        \includegraphics[width=\textwidth]{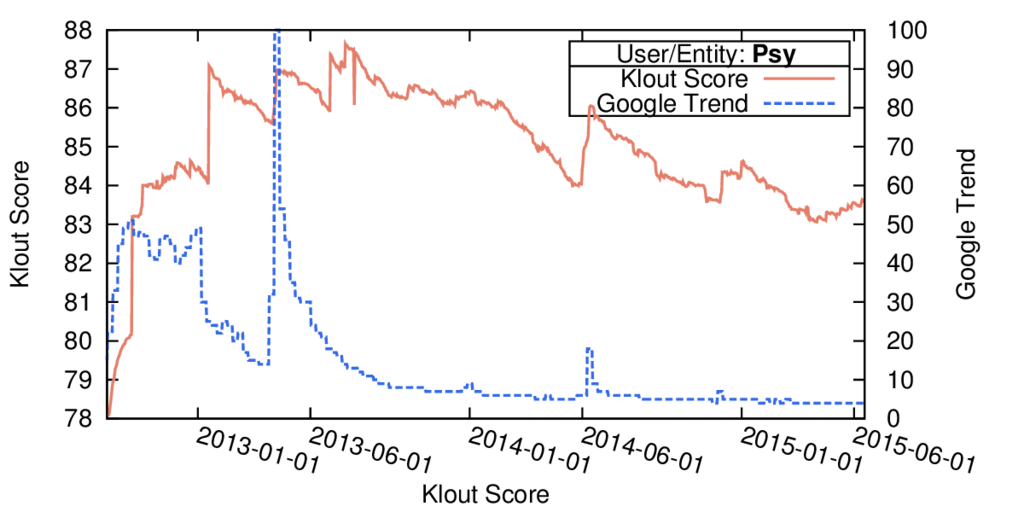}
\end{subfigure}
\begin{subfigure}[b]{0.4\textwidth}
        \includegraphics[width=\textwidth]{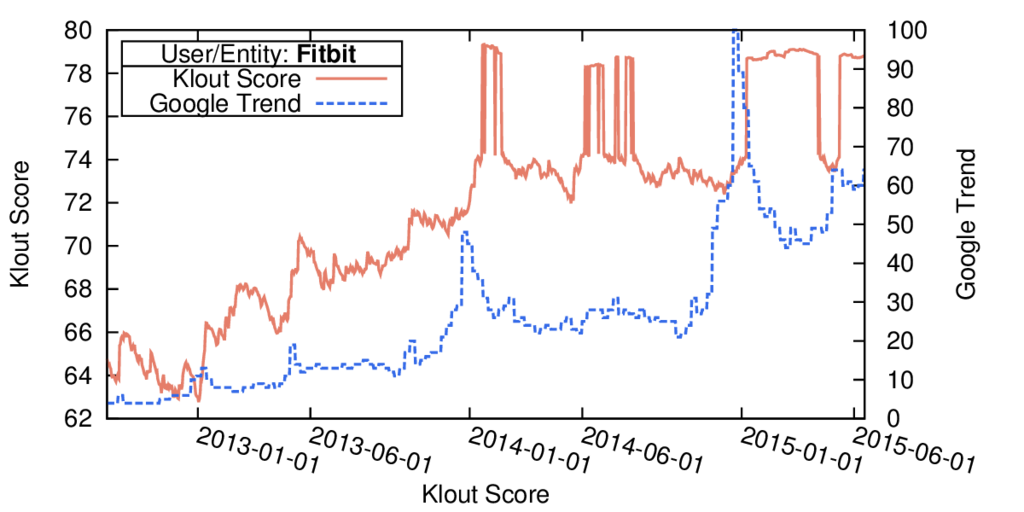}
\end{subfigure}
\caption{Comparison of Klout Score and Google Trends}
\label{fig:google_trend}
\end{figure*}

\subsection{Spread of Information}

To validate the effectiveness of the Klout Score, we ran a year long experiment to measure the spread of information with respect to the user scores. 
Users with Klout Scores varying in the range of $[10-80]$ were targeted with \textit{perks}, which could be claimed by the users. 
The users were encouraged, but not mandated, to post messages about their experience with the claimed perk.
The users' audiences then reacted to these messages, with a higher number of reactions indicating a greater spread of information.
$87,675$ users posted messages after claiming a perk, out of which $18,308$ posts recieved a total of $394,083$ reactions.

The average number of reactions are plotted on a log scale against the targeted users' Klout scores in Figure \ref{figure:perk_analysis}.
The curve shows a monotonically increasing curve, where users with higher scores show a higher number of reactions.
We also observe an order of magnitude difference in reactions recieved by users with a score of 60 compared to a score of 30, and similarly for users with a score of 80 compared to 60.
This validates that users with higher Klout scores are able to spread information more effectively in a network.

\subsection{Comparisons with Other Systems}

\begin{table}[htdp]
\caption{Comparison with ATP Tennis Player Ranking and Forbes Most Powerful Women Ranking}
\label{tbl:comparison_ATP_Forbes} 
  \centering
  \footnotesize  
  \tabcolsep=0.11cm
    \begin{tabular}{|c||c|c||c|c|}
    \hline
    \rowcolor{Gray} $Ranking$ & ATP & Klout & Forbes & Klout\\ \hline
      1  &  Novak Djokovic  & 89.54   & Hillary Clinton   &  93.23\\ \hline
      2  &  Roger Federer   & 90.26   & Melinda Gates     &  83.57\\ \hline
      3  &  Andy Murray     & 89.50   & Mary Barra        &  77.53\\ \hline
      4  &  Stan Wawrinka   & 86.86   & Christine Lagarde &  83.89\\ \hline
      5  &  Kei Nishikori   & 83.50   & Dilma Rousseff    &  86.84\\ \hline
      6  &  Tomas Berdych   & 66.69   & Sheryl Sandberg   &  83.18\\ \hline
      7  &  David Ferrer    & 65.98   & Susan Wojcicki    &  80.04\\ \hline
      8  &  Milos Raonic    & 82.28   & Michelle Obama    &  87.30\\ \hline
      9  &  Marin Cilic     & 58.93   & Park Geun-hye     &  81.80\\ \hline
      10 &  Rafael Nadal    & 82.37   & Oprah Winfrey     &  91.08\\ \hline
    \end{tabular}
\end{table}

\subsubsection{Real-World Rankings}

We also compare the Klout Score with other real world rankings that indicate influence.
Table \ref{tbl:comparison_ATP_Forbes} shows the Klout scores compared with ATP rankings for tennis players\footnote{http://www.atpworldtour.com/en/rankings/singles}, and Forbes' list for most powerful women\footnote{http://www.forbes.com/power-women/}, as of June 2015.

To measure the ranking quality of Klout Score, we adopt the \textit{normalized Discounted Cumulative Gain} (nDCG) metric, defined in Eq. \ref{eq:NDCG}. 
The \textit{Discounted Cumulative Gain} upto position $p$ (DCG$_p$) is calculated as Eq.\ref{eq:DCG}, and the ideal DCG for $p$ is denoted by IDCG$_p$.
 
\begin{equation}
nDCG_p = \dfrac{DCG_p}{IDCG_p}
\label{eq:NDCG}
\end{equation}

\begin{equation}
DCG_p = \sum\limits_{i=1}^{p} \dfrac{2^{rel_i}-1}{log_2{(i+1)}}
\label{eq:DCG}
\end{equation}

We calculate the IDCG$_p$ by using the ATP or Forbes rankings as the ideal ordering of users.
We set the relevance $rel$ of a person as $p / rank_{ideal}$, where the $rank_{ideal}$ is her/his position in the ideal ranking.
For example, for $p=10$ the relevance of \textit{Novak Djokovic} is $10$ because his position is $1$ in the ATP ranking.
Thus his contribution to the IDCG$_p$ measure is $\frac{2^{10} - 1}{log_2{2}}$, whereas his contribution to the DCG$_p$ measure for the Klout Score ranking is $\frac{2^{10} - 1}{log_2{3}}$, since he appears in the 2nd position there.
Setting the relevance in this manner places stronger emphasis in retrieving correct higher ranked documents.

With this setting, the nDCG$_{10}$ measure for the Klout score with respect to the ATP and Forbes ranking is computed as $0.878$ and $0.874$, respectively.
This demonstrates that the Klout score is able to capture real world influence to a high degree for these examples.

\subsubsection{Google Trends}
To observe temporal sensitivity, we plot the Klout Score for a few entities along with their Google trends for the last three years in Figure \ref{fig:google_trend}.
For both \textit{Starbucks} and \textit{Airbnb}, the Klout scores show similar fluctuations compared to their Google Trends, indicating a strong correlation between online influence and search popularity.
For \textit{Fitbit} the Klout score catches a few spikes that are not seen in Google Trends.
This also reveals that the Klout score is sensitive to short-term variations, and tracks such changes very closely. 

For the music artist \textit{Psy}, we see that the Google Trend drops significantly after 2013-06-01 while his Klout score decreases more gradually. 
This is because while the Google Trend reflects only the immediate short-term popularity, the Klout score incorporates long-lasting features as well.

\begin{figure*}
  \centering
  \includegraphics[width=\textwidth]{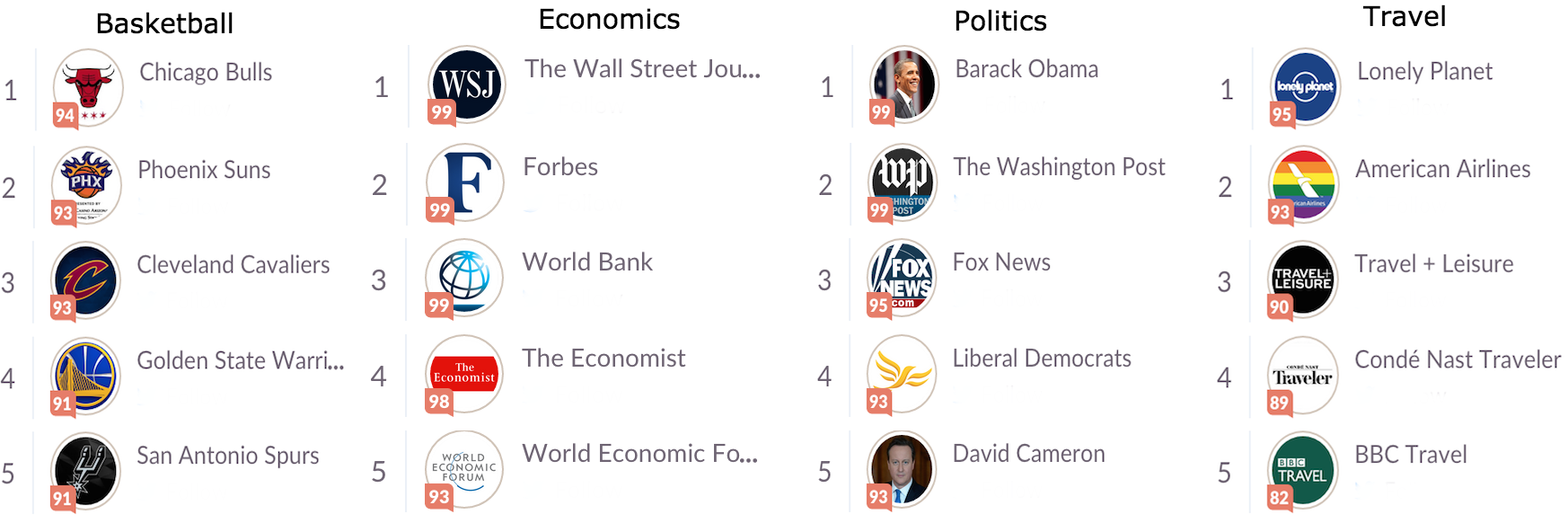}
  \caption{Top-5 Influencers by Topic}
  \label{fig:topic_influencer}
\end{figure*}

\subsection{Influencers By Topic}
Since influence is typically contextual, we explore the effectiveness of the Klout score across different topical domains. 
Users in topical domains were identified using the methodology described in \cite{nemanja-lasta}, and ranked by their Klout scores within their respective domains.
Figure \ref{fig:topic_influencer} shows the top $5$ ranked influencers in selected topics.
For a topic such as politics, we see that the highest scored users includes prominent politicians such as are \textit{Barack Obama} and \textit{David Cameron}, as well as news and media entites such as \textit{The Washington Post} and \textit{Fox News}, all of whom are clearly very influential.
These examples clearly depict that Klout score can correctly identify the influencers in a variety of domains.


\section{Conclusion}
\label{section:conclusion}
In this work, we propose a hierarchical scoring system called the Klout Score, that assigns influence scores to 750 million users across 9 different social networks, by analyzing 45 billion interactions daily.
The framework scales to hundreds of millions of users by leveraging distributed batch processing frameworks that aggregate signals in linear time with respect to the nodes in the graph.

To create the score, a feature generation framework aggregates signals across several dimensions for each user, creating a large feature set containing over two thousand features. 
In addition to incoporating signals from social network interactions, the features also incorporate factors such as Wikipedia that provide a proxy for real world influence. 
Weights obtained from supervised models are applied to these features to generate network or community scores.
These scores are then combined hierarchically to get the final Klout Score.


Our experiments also validate that users with higher Klout Scores are able to spread information wider in a network.
We further compare the performance of the score against other ranking systems and also analyze the dynamic nature of the score. 
We examine different topical domains and find that highly influential users are correctly identified within these domains by their high scores.

The Klout Score presented here is, of course, only a partial representation of the influence of a user.
Nevertheless, by building an extensible feature generation framework and a hierarchical scoring structure, the system is able to easily incorporate new sources of information, and therefore grow more accurate over time.
Several applications may be potentially built using an influence scoring system such as the Klout Score, and we hope this work enables future work in this area.

\section{Acknowledgement}
\label{section:acknowledgement}
We would like to thank our former colleagues Jerome Banks, Girish Lingappa, Andras Benke, Alexy Khrabrov and Ding Zhou who contributed towards building the Klout Score. We also thank Joe Fernandez for his valuable ideas and feedback throughout the project. 


\bibliographystyle{IEEEtran}
\bibliography{klout_score_abstract}

\begin{thebibliography}{10}
\providecommand{\url}[1]{#1}
\csname url@samestyle\endcsname
\providecommand{\newblock}{\relax}
\providecommand{\bibinfo}[2]{#2}
\providecommand{\BIBentrySTDinterwordspacing}{\spaceskip=0pt\relax}
\providecommand{\BIBentryALTinterwordstretchfactor}{4}
\providecommand{\BIBentryALTinterwordspacing}{\spaceskip=\fontdimen2\font plus
\BIBentryALTinterwordstretchfactor\fontdimen3\font minus
  \fontdimen4\font\relax}
\providecommand{\BIBforeignlanguage}[2]{{%
\expandafter\ifx\csname l@#1\endcsname\relax
\typeout{** WARNING: IEEEtran.bst: No hyphenation pattern has been}%
\typeout{** loaded for the language `#1'. Using the pattern for}%
\typeout{** the default language instead.}%
\else
\language=\csname l@#1\endcsname
\fi
#2}}
\providecommand{\BIBdecl}{\relax}
\BIBdecl

\bibitem{anger2011klout}
I.~Anger and C.~Kittl, ``Measuring influence on twitter,'' in \emph{i-KNOW
  '11}, 2011.

\bibitem{schaefer2012influence}
S.~M., ``Return on influence: The revolutionary power of klout, social scoring,
  and influence marketing.'' in \emph{McGraw-Hill Professional: London}, 2012.

\bibitem{tang2009social}
J.~Tang, J.~Sun, C.~Wang, and Z.~Yang, ``Social influence analysis in
  large-scale networks,'' in \emph{SIGKDD'09}, 2009.

\bibitem{agarwal2008identifying}
N.~Agarwal, H.~Liu, L.~Tang, and P.~S. Yu, ``Identifying the influential
  bloggers in a community,'' in \emph{WSDM'08}, 2008.

\bibitem{weng2010twitterrank}
J.~Weng, E.-P. Lim, J.~Jiang, and Q.~He, ``Twitterrank: finding topic-sensitive
  influential twitterers,'' in \emph{WSDM'10}, 2010.

\bibitem{Kempe2003MSI}
D.~Kempe, J.~Kleinberg, and E.~Tardos, ``Maximizing the spread of influence
  through a social network,'' in \emph{SIGKDD'03}, 2003.

\bibitem{Chen2009EIM}
W.~Chen, Y.~Wang, and S.~Yang, ``Efficient influence maximization in social
  networks,'' in \emph{SIGKDD'09}, 2009.

\bibitem{chen2015online}
S.~Chen, J.~Fan, G.~Li, J.~Feng, K.-l. Tan, and J.~Tang, ``Online topic-aware
  influence maximization,'' \emph{Proceedings of the VLDB Endowment}, vol.~8,
  no.~6, pp. 666--677, 2015.

\bibitem{khrabrov2010discovering}
A.~Khrabrov and G.~Cybenko, ``Discovering influence in communication networks
  using dynamic graph analysis,'' in \emph{IEEE Second International Conference
  on Social Computing (SocialCom)}, 2010.

\bibitem{hajian2011modelling}
B.~Hajian and T.~White, ``Modelling influence in a social network: Metrics and
  evaluation,'' in \emph{IEEE Third Inernational Conference on Social Computing
  (SocialCom)}, 2011.

\bibitem{Spasojevic:when-to-post}
N.~Spasojevic, Z.~Li, A.~Rao, and P.~Bhattacharyya, ``When-to-post on social
  networks,'' in \emph{Proc. of ACM Conference on Knowledge Discovery and Data
  Mining (KDD)}, ser. KDD '15, 2015.

\bibitem{edwards2013credibility}
C.~Edwards, P.~R. Spence, C.~J. Gentile, A.~Edwards, and A.~Edwards, ``How much
  klout do you have … a test of system generated cues on source
  credibility.'' vol.~29, 2013, p. A12–A16.

\bibitem{nemanja-lasta}
N.~Spasojevic, J.~Yan, A.~Rao, and P.~Bhattacharyya, ``Lasta: Large scale topic
  assignment on multiple social networks,'' in \emph{Proc. of ACM Conference on
  Knowledge Discovery and Data Mining (KDD)}, ser. KDD '14, 2014.

\end{thebibliography}

\end{document}